\documentclass[10pt]{article}
\usepackage[OE]{express}
\usepackage[english]{babel}
\usepackage[utf8]{inputenc}
\begin{document}

\title{Dynamics of Ultrafast Laser Ablation of Water}

\author{Javier Hernandez-Rueda\authormark{1,2} and Dries van Oosten\authormark{1,3}}

\address{\authormark{1}Debye Institute for Nanomaterials Science and Center for Extreme Matter and Emergent Phenomena, Utrecht University, Princetonplein 5, 3584 CC Utrecht, The Netherlands\\
\authormark{2}j.hernandezrueda@uu.nl\\
\authormark{3}d.vanoosten@uu.nl}




\begin{abstract}
Ultrafast laser ablation is an extremely precise and clean method of removing material, applied in material processing as well as medical applications. And due to its violent nature, it tests our understanding of the interplay between optics, condensed matter physics and fluid dynamics. In this manuscript, we experimentally investigate the femtosecond laser induced explosive vaporization of water at a water/gas interface on the micron-scale through several time-scales. Using time-resolved microscopy in reflection mode, we observe the formation of a hot electron plasma, an explosively expanding water vapor and a shockwave propelled into the surrounding gas. We study this fs-laser induced water vapor expansion dynamics in the presence of different atmospheres, i.e. Helium, air and tetrafluoroethane. We use the Sedov-Taylor model to explain the expansion of the water vapor and estimate the energy released in the process. 
\end{abstract}



\bibliographystyle{unsrt}
\bibliography{LibraryOL2018}

\begin{thebibliography}{10}

\bibitem{Stan2016}
Claudiu~A. Stan, Despina Milathianaki, Hartawan Laksmono, Raymond~G. Sierra,
  Trevor~A. McQueen, Marc Messerschmidt, Garth~J. Williams, Jason~E. Koglin,
  Thomas~J. Lane, Matt~J. Hayes, Serge A.~H. Guillet, Mengning Liang, Andrew~L.
  Aquila, Philip~R. Willmott, Joseph~S. Robinson, Karl~L. Gumerlock, Sabine
  Botha, Karol Nass, Ilme Schlichting, Robert~L. Shoeman, Howard~A. Stone, and
  Sébastien Boutet.
\newblock Liquid explosions induced by x-ray laser pulses.
\newblock {\em Nature Physics}, 12:966, May 2016.

\bibitem{silberberg2001laser}
Yaron Silberberg.
\newblock Laser science: Physics at the attosecond frontier.
\newblock {\em Nature}, 414(6863):494, 2001.

\bibitem{Schaffer2002}
Chris~B. Schaffer, Nozomi Nishimura, Eli~N. Glezer, Albert M.-T. Kim, and Eric
  Mazur.
\newblock Dynamics of femtosecond laser-induced breakdown in water from
  femtoseconds to microseconds.
\newblock {\em Opt. Express}, 10(3):196--203, Feb 2002.

\bibitem{Strycker2013}
B.D. Strycker, M.M. Springer, A.J. Traverso, A.A. Kolomenskii, G.W. Kattawar,
  and A.V. Sokolov.
\newblock Femtosecond-laser-induced shockwaves in water generated at an
  air-water interface.
\newblock {\em Opt. Express}, 21(20):23772--23784, Oct 2013.

\bibitem{zhang2015modeling}
Hao Zhang, SA~Wolbers, DM~Krol, JI~Dijkhuis, and D~Van~Oosten.
\newblock Modeling and experiments of self-reflectivity under femtosecond
  ablation conditions.
\newblock {\em JOSA B}, 32(4):606--616, 2015.

\bibitem{Gattass2008}
Rafael~R. Gattass and Eric Mazur.
\newblock Femtosecond laser micromachining in transparent materials.
\newblock {\em Nature Photonics}, 2:219, April 2008.

\bibitem{zhang2013self}
Hao Zhang, Denise~M Krol, Jaap~I Dijkhuis, and Dries van Oosten.
\newblock Self-scattering effects in femtosecond laser nanoablation.
\newblock {\em Optics letters}, 38(23):5032--5035, 2013.

\bibitem{hernandez2017influence}
Javier Hernandez-Rueda, Jasper Clarijs, Dries van Oosten, and Denise~M Krol.
\newblock The influence of femtosecond laser wavelength on waveguide
  fabrication inside fused silica.
\newblock {\em Applied Physics Letters}, 110(16):161109, 2017.

\bibitem{Tirlapur2002}
Uday~K. Tirlapur and Karsten König.
\newblock Targeted transfection by femtosecond laser.
\newblock {\em Nature}, 418:290, July 2002.

\bibitem{Linz2015}
Norbert Linz, Sebastian Freidank, Xiao-Xuan Liang, Hannes Vogelmann, Thomas
  Trickl, and Alfred Vogel.
\newblock Wavelength dependence of nanosecond infrared laser-induced breakdown
  in water: Evidence for multiphoton initiation via an intermediate state.
\newblock {\em Phys. Rev. B}, 91:134114, Apr 2015.

\bibitem{Collins2008}
Hazel~A. Collins, Mamta Khurana, Eduardo~H. Moriyama, Adrian Mariampillai, Emma
  Dahlstedt, Milan Balaz, Marina~K. Kuimova, Mikhail Drobizhev, Victor X.~D.
  Yang, David Phillips, Aleksander Rebane, Brian~C. Wilson, and Harry~L.
  Anderson.
\newblock Blood-vessel closure using photosensitizers engineered for two-photon
  excitation.
\newblock {\em Nature Photonics}, 2:420, May 2008.

\bibitem{ditmire1999nuclear}
Todd Ditmire, J~Zweiback, VP~Yanovsky, TE~Cowan, G~Hays, and KB~Wharton.
\newblock Nuclear fusion from explosions of femtosecond laser-heated deuterium
  clusters.
\newblock {\em Nature}, 398(6727):489--492, 1999.

\bibitem{hentschel2001attosecond}
M~Hentschel, R~Kienberger, Ch~Spielmann, Georg~A Reider, N~Milosevic, Thomas
  Brabec, Paul Corkum, Ulrich Heinzmann, Markus Drescher, and Ferenc Krausz.
\newblock Attosecond metrology.
\newblock {\em Nature}, 414(6863):509, 2001.

\bibitem{Yanik2004}
Mehmet~Fatih Yanik, Hulusi Cinar, Hediye~Nese Cinar, Andrew~D. Chisholm, Yishi
  Jin, and Adela Ben-Yakar.
\newblock Functional regeneration after laser axotomy.
\newblock {\em Nature}, 432:822, December 2004.

\bibitem{Guo2008}
Samuel~X. Guo, Frederic Bourgeois, Trushal Chokshi, Nicholas~J. Durr,
  Massimo~A. Hilliard, Nikos Chronis, and Adela Ben-Yakar.
\newblock Femtosecond laser nanoaxotomy lab-on-a-chip for in vivo nerve
  regeneration studies.
\newblock {\em Nature Methods}, 5:531, April 2008.

\bibitem{Vogel2003}
Alfred Vogel and Vasan Venugopalan.
\newblock Mechanisms of pulsed laser ablation of biological tissues.
\newblock {\em Chemical Reviews}, 103(2):577--644, 2003.
\newblock PMID: 12580643.

\bibitem{Williamson1997}
J.~Charles Williamson, Jianming Cao, Hyotcherl Ihee, Hans Frey, and Ahmed~H.
  Zewail.
\newblock Clocking transient chemical changes by ultrafast electron
  diffraction.
\newblock {\em Nature}, 386:159, March 1997.

\bibitem{sundaram2002inducing}
SK~Sundaram and E~Mazur.
\newblock Inducing and probing non-thermal transitions in semiconductors using
  femtosecond laser pulses.
\newblock {\em Nature Materials}, 1:217--224, 2002.

\bibitem{PhysRevLett.81.224}
K.~Sokolowski-Tinten, J.~Bialkowski, A.~Cavalleri, D.~von~der Linde, A.~Oparin,
  J.~Meyer-ter Vehn, and S.~I. Anisimov.
\newblock Transient states of matter during short pulse laser ablation.
\newblock {\em Phys. Rev. Lett.}, 81:224--227, Jul 1998.

\bibitem{douhal1995femtosecond}
A~Douhal, SK~Kim, and AH~Zewail.
\newblock Femtosecond molecular dynamics of tautomerization in model base
  pairs.
\newblock {\em Nature}, 378(6554):260, 1995.

\bibitem{Linde1997}
D~von~der Linde, K~Sokolowski-Tinten, and J~Bialkowski.
\newblock Laser–solid interaction in the femtosecond time regime.
\newblock {\em Applied Surface Science}, 109-110:1 -- 10, 1997.

\bibitem{Siders1999}
C.~W. Siders, A.~Cavalleri, K.~Sokolowski-Tinten, Cs. Toth, T.~Guo, M.~Kammler,
  M.~Horn von Hoegen, K.~R. Wilson, D.~von~der Linde, and C.~P.~J. Barty.
\newblock Detection of nonthermal melting by ultrafast x-ray diffraction.
\newblock {\em Science}, 286:1340, November 1999.

\bibitem{Temnov:06}
Vasily~V. Temnov, Klaus Sokolowski-Tinten, Ping Zhou, and Dietrich von~der
  Linde.
\newblock Ultrafast imaging interferometry at femtosecond-laser-excited
  surfaces.
\newblock {\em J. Opt. Soc. Am. B}, 23(9):1954--1964, Sep 2006.

\bibitem{2009Nagy}
Filkorn T Sarayba~M. Nagy~Z, Takacs~A.
\newblock Initial clinical evaluation of an intraocular femtosecond laser in
  cataract surgery.
\newblock {\em J Refract Surg. 2009; 25}, 2009.

\bibitem{Sarpe2012}
C~Sarpe, J~Köhler, T~Winkler, M~Wollenhaupt, and T~Baumert.
\newblock Real-time observation of transient electron density in water
  irradiated with tailored femtosecond laser pulses.
\newblock {\em New Journal of Physics}, 14(7):075021, 2012.

\bibitem{Krueger1993}
Ronald~R. Krueger, Jerzy~S. Krasinski, Czeslaw Radzewicz, Karl~G. Stonecipher,
  and J.~James Rowsey.
\newblock Photography of shock waves during excimer laser ablation of the
  cornea: Effect of helium gas on propagation velocity.
\newblock {\em Cornea}, 12(4):330--334, 1993.

\bibitem{Juhasz1994}
Tibor Juhasz, Xin~H. Hu, Laszlo Turi, and Zsolt Bor.
\newblock Dynamics of shock waves and cavitation bubbles generated by
  picosecond laser pulses in corneal tissue and water.
\newblock {\em Lasers in Surgery and Medicine}, 15(1):91--98, 1994.

\bibitem{Sarpe-Tudoran2006}
C.~Sarpe-Tudoran, A.~Assion, M.~Wollenhaupt, M.~Winter, and T.~Baumert.
\newblock Plasma dynamics of water breakdown at a water surface induced by
  femtosecond laser pulses.
\newblock {\em Applied Physics Letters}, 88(26):261109, 2006.

\bibitem{Rethfeld2004a}
B.~Rethfeld.
\newblock Unified model for the free-electron avalanche in laser-irradiated
  dielectrics.
\newblock {\em Phys. Rev. Lett.}, 92:187401, May 2004.

\bibitem{Rethfeld2006}
B.~Rethfeld.
\newblock Free-electron generation in laser-irradiated dielectrics.
\newblock {\em Phys. Rev. B}, 73:035101, Jan 2006.

\bibitem{Vogel1996}
A.~Vogel, S.~Busch, and U.~Parlitz.
\newblock Shock wave emission and cavitation bubble generation by picosecond
  and nanosecond optical breakdown in water.
\newblock {\em The Journal of the Acoustical Society of America},
  100(1):148--165, 1996.

\bibitem{Apitz2005}
I.~Apitz and A.~Vogel.
\newblock Material ejection in nanosecond er:yag laser ablation of water,
  liver, and skin.
\newblock {\em Applied Physics A}, 81(2):329--338, Jul 2005.

\bibitem{Zeng2006}
Xianzhong Zeng, Xianglei Mao, Samuel~S. Mao, Sy-Bor Wen, Ralph Greif, and
  Richard~E. Russo.
\newblock Laser-induced shockwave propagation from ablation in a cavity.
\newblock {\em Applied Physics Letters}, 88(6):061502, 2006.

\bibitem{thoroddsen2009spray}
Sigurdur~T Thoroddsen, K~Takehara, TG~Etoh, and C-D Ohl.
\newblock Spray and microjets produced by focusing a laser pulse into a
  hemispherical drop.
\newblock {\em Physics of Fluids}, 21(11):112101, 2009.

\bibitem{tagawa2012highly}
Yoshiyuki Tagawa, Nikolai Oudalov, Claas~Willem Visser, Ivo~R Peters, Devaraj
  van~der Meer, Chao Sun, Andrea Prosperetti, and Detlef Lohse.
\newblock Highly focused supersonic microjets.
\newblock {\em Physical review X}, 2(3):031002, 2012.

\bibitem{klein2015drop}
Alexander~L Klein, Wilco Bouwhuis, Claas~Willem Visser, Henri Lhuissier, Chao
  Sun, Jacco~H Snoeijer, Emmanuel Villermaux, Detlef Lohse, and Hanneke
  Gelderblom.
\newblock Drop shaping by laser-pulse impact.
\newblock {\em Physical review applied}, 3(4):044018, 2015.

\bibitem{avila2016fragmentation}
Silvestre Roberto~Gonzalez Avila and Claus-Dieter Ohl.
\newblock Fragmentation of acoustically levitating droplets by laser-induced
  cavitation bubbles.
\newblock {\em Journal of Fluid Mechanics}, 805:551--576, 2016.

\bibitem{Nguyen2018}
Thao~T.P. Nguyen, Rie Tanabe, and Yoshiro Ito.
\newblock Comparative study of the expansion dynamics of laser-driven plasma
  and shock wave in in-air and underwater ablation regimes.
\newblock {\em Optics \& Laser Technology}, 100:21 -- 26, 2018.

\bibitem{vogel2005mechanisms}
Alfred Vogel, J~Noack, G~H{\"u}ttman, and G~Paltauf.
\newblock Mechanisms of femtosecond laser nanosurgery of cells and tissues.
\newblock {\em Applied Physics B}, 81(8):1015--1047, 2005.

\bibitem{liu1982simple}
JM~Liu.
\newblock Simple technique for measurements of pulsed gaussian-beam spot sizes.
\newblock {\em Optics letters}, 7(5):196--198, 1982.

\bibitem{keldysh1965ionization}
LV~Keldysh et~al.
\newblock Ionization in the field of a strong electromagnetic wave.
\newblock {\em Sov. Phys. JETP}, 20(5):1307--1314, 1965.

\bibitem{Hernandez-Rueda2012}
J.~Hernandez-Rueda, D.~Puerto, J.~Siegel, M.~Galvan-Sosa, and J.~Solis.
\newblock Plasma dynamics and structural modifications induced by femtosecond
  laser pulses in quartz.
\newblock {\em Applied Surface Science}, 258(23):9389 -- 9393, 2012.
\newblock EMRS 2011 Spring Symp J: Laser Materials Processing for Micro and
  Nano Applications.

\bibitem{Hernandez-Rueda2015}
J.~Hernandez-Rueda, J.~Siegel, M.~Galvan-Sosa, A.~Ruiz de~la Cruz,
  M.~Garcia-Lechuga, and J.~Solis.
\newblock Controlling ablation mechanisms in sapphire by tuning the temporal
  shape of femtosecond laser pulses.
\newblock {\em J. Opt. Soc. Am. B}, 32(1):150--156, Jan 2015.

\bibitem{Arnold1999}
N.~Arnold, J.~Gruber, and J.~Heitz.
\newblock Spherical expansion of the vapor plume into ambient gas: an
  analytical model.
\newblock {\em Applied Physics A}, 69(1):S87--S93, Dec 1999.

\bibitem{mafune2000formation}
Fumitaka Mafun{\'e}, Jun-ya Kohno, Yoshihiro Takeda, Tamotsu Kondow, and
  Hisahiro Sawabe.
\newblock Formation and size control of silver nanoparticles by laser ablation
  in aqueous solution.
\newblock {\em The Journal of Physical Chemistry B}, 104(39):9111--9117, 2000.

\bibitem{wagner2010euv}
Christian Wagner and Noreen Harned.
\newblock Euv lithography: Lithography gets extreme.
\newblock {\em Nature Photonics}, 4(1):24, 2010.

\end{thebibliography}

\section{Introduction}

In many applications in research \cite{Stan2016, silberberg2001laser, Schaffer2002, Strycker2013,zhang2015modeling}, technology \cite{Gattass2008, zhang2013self,hernandez2017influence} and healthcare \cite{Tirlapur2002, Linz2015, Collins2008}, ultrafast lasers are employed to cut and remove material as well as to locally modify the chemical, structural and optical properties of the target. In even more extreme examples, pulsed lasers are used to trigger nuclear fusion \cite{ditmire1999nuclear} and to generate high harmonics in attosecond science \cite{hentschel2001attosecond} or more practically, to study nerve regeneration $\it{in}$ $\it{vivo}$ after fs-laser axotomy \cite{Yanik2004, Guo2008}. The unprecedented spatial resolution achieved in all these studies is partly due to the simultaneous ultrafast character and non-linear nature of the light-matter interaction, which leads to a reduced heat affected zone and to minimal collateral damage when compared to the outcome achieved using longer pulses \cite{Vogel2003}. These outstanding features allow researchers not only to accurately modify materials but also to investigate the fast and ultrafast mechanisms involved during the process of modification, such as phase transitions or bio-chemical and chemical reactions \cite{Williamson1997, sundaram2002inducing, Stan2016, PhysRevLett.81.224}. In this way, ultrashort laser pulses, with a duration that ranges from a few to hundreds of femtoseconds (1 fs = 10$^{-15}$ s), can be used to probe the ultrafast chain of processes triggered by an excitation laser pulse in a so-called pump and probe system \cite{douhal1995femtosecond, Williamson1997, PhysRevLett.81.224, Linde1997, Siders1999, Temnov:06}. 

Particularly spectacular and very relevant for biological applications and laser-based surgery, is the ablation of aqueous media \cite{Schaffer2002, Vogel2003, 2009Nagy, Tirlapur2002, Guo2008}. Crucial steps in this inherently multi-scale process are the absorption of laser-energy by the aqueous medium, resulting in local heating, followed by the evaporation of the liquid, which in turn does work against the tissue in which it is embedded \cite{Vogel2003, 2009Nagy, Sarpe2012, Krueger1993, Juhasz1994, Linz2015, Stan2016, Sarpe-Tudoran2006, Strycker2013}. The use of fs-laser pulses is particularly interesting for such applications, as the absorption in that case is very nonlinear and therefore is limited to a region that is typically smaller than the focal volume \cite{Siders1999, Rethfeld2004a, PhysRevLett.81.224}. The extreme optical nonlinearity makes the description of the absorption of the ultra-short laser pulse very challenging, but on the other hand, leads to a very attractive separation of timescales \cite{Rethfeld2004a, Rethfeld2006}. In the first picosecond, a hot electron plasma is created. Within the first 10 picoseconds, thus after the laser pulse is gone, this plasma equilibrates in temperature with the surrounding liquid. As a result, the liquid becomes superheated and explosively evaporates. The interaction dynamics of pulsed lasers with water has been investigated using lateral imaging via time-resolved shadowgraphy or time-resolved scatterometry, for instance looking at the propagation of laser-induced shockwaves and cavitation bubbles \cite{Schaffer2002, Strycker2013, Vogel1996, Apitz2005, Zeng2006, thoroddsen2009spray, tagawa2012highly, klein2015drop, avila2016fragmentation, Nguyen2018}. Lateral imaging provides a wealth of information of the aftermath for long time-delays (ns-$\mu$s), at the expense of losing lateral spatial resolution to study the microscopic features of the initial electron plasma under strong focusing conditions \cite{Schaffer2002}, relevant for instance to fs-laser cell surgery and neurosurgery \cite{Tirlapur2002, vogel2005mechanisms}.

In this work, we study femtosecond laser ablation at a water/gas interface using time-resolved imaging with submicron optical resolution in reflection mode, from 10 femtosecond to 100 nanosecond time-scales. The experiment is carried out inside three different gaseous atmospheres with increasing molecular weights and densities at room temperature, $\it{i.e.}$ Helium (4 g/mol), air (18 g/mol) and tetrafluoroethane (102 g/mol). The already mentioned separation of time-regimes allows us to compartmentalize the multi-scale problem and separately observe the light-water interaction (fs-ps), the evaporation process (10 ps) and the supersonic vapor expansion dynamics (ns). We interpret our experimental observations with a Sedov-Taylor model to extract an estimate of the energy carried by the expanding vapor. 

\section{Experimental procedure}

\subsection{Experimental setup}

Fig.1 (a) shows a detailed scheme of the collinear pump and probe setup. The laser source used during the experiments is a femtosecond regenerative amplifier (Hurricane, Spectra-Physics) that produces 150~fs laser pulses at a wavelength of 800~nm. Using a $\lambda$/2 waveplate and a polarized beam cube, we split the 800~nm laser pulse into two sub-pulses. The most energetic pulse runs over a fixed delay line to the microscope objective while the weaker fraction runs over an automated delay line to tune the pump-probe delay time up to a maximum of $1.4~$ns (Newport Co.). For larger delays, we introduce longer detours in the probe optical path to obtain delays up to $100~$ns. Before the delay line, a BBO (beta barium borate) crystal is used to frequency double the probe pulse and an edge filter (F) is used to block the remaining 800~nm light. The 400~nm probe light is then spatially and temporally overlapped with the 800~nm pump light in a pellicle beam splitter (PBS) before both collinearly enter an infinity corrected microscope objective (Nikon CFI60, 100X, NA = 0.8). Additional lenses in both optical paths are used to ensure that the 800~nm pump (one-to-one telescope, T0) is strongly focused by the objective, while the 400 nm probe is focused (L) in the back focal plane of the objective, resulting in wide-field illumination suitable for imaging the surface of the sample over an area of $45\times45~\mu$m$^2$ onto the EMCCD camera chip (Andor, iXon 885).

\begin{figure}[ht]
\includegraphics[width=13cm]{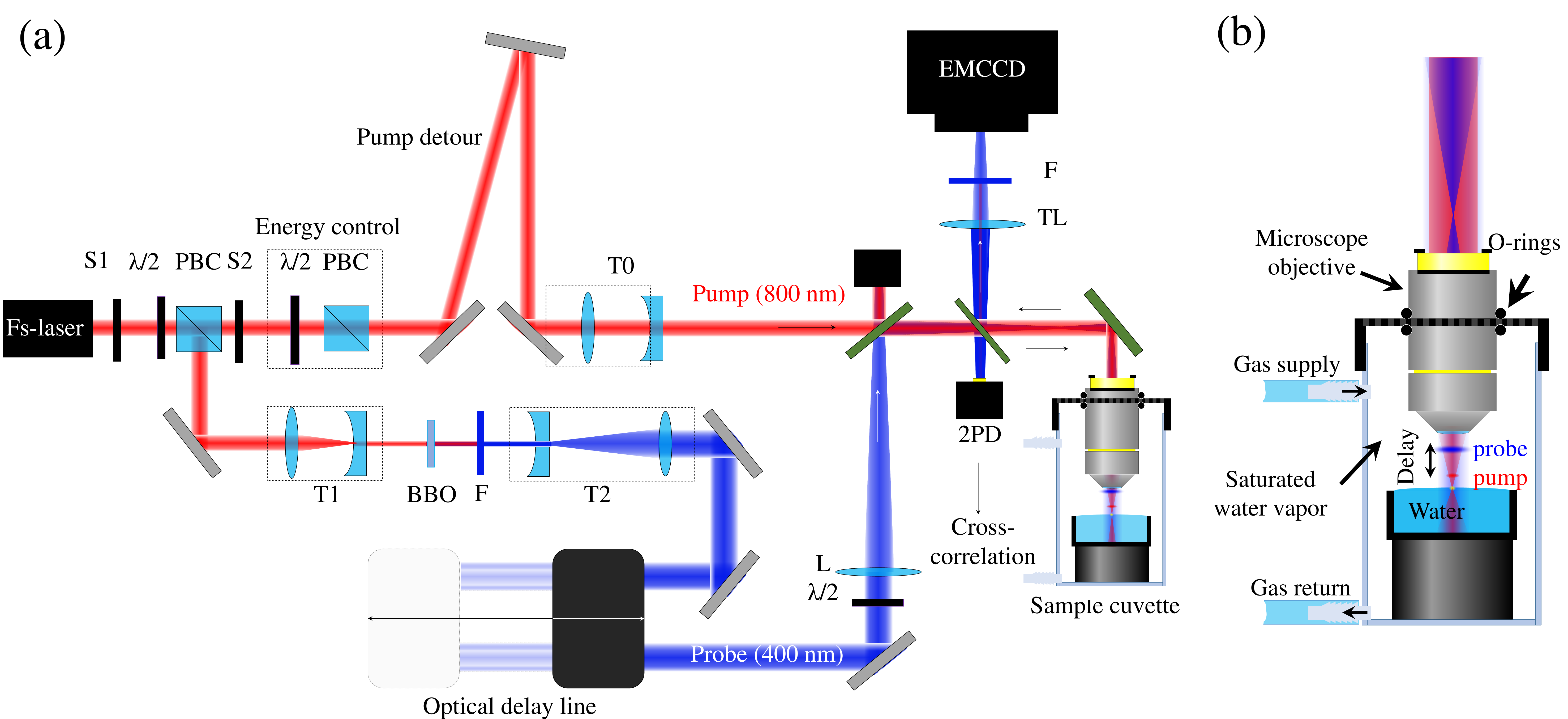}
\caption{(a) Schematic of the femtosecond time-resolved ablation setup. (b) Detail of the sample cuvette.}
\label{fig1}
\end{figure}

The reflected light is collected by the objective and used to image the water/air interface through a 300~mm tube lens (TL) as shown in Fig.1 (a). A filter (F) that only transmits the 400~nm light ($\Delta \lambda_{\rm FWHM}=10~$nm, $\lambda_{\rm 0}=400~$nm) is used to prevent 800~nm pump light and plasma emission from reaching the camera. The optical path between the tube lens/filter and the camera chip is tubed in order to minimize the collection of stray ambient light. The delay control unit of the Pockels cells is synchronized with two optomechanical shutters (S1, S2) to ensure single shot experiments either with the probe pulse only (S1 open) or with both pump and probe pulses (both S1 and S2 open). For each temporal delay, several images are taken, each corresponding to a single shot of the laser while both pump and probe pulses are present. Additionally, several reference images are acquired. It is worth noting that unlike in the case of solid targets, the water surface self-restores a few milliseconds after strong laser excitation. This prevents incubation effects and allows us to record several pump-probe images at the same spot without re-focusing or moving the sample to a fresh area. The beam waist (1/e$^2$) of the focused pump beam at the sample surface was calibrated to be $3.1~\mu$m using Liu method \cite{liu1982simple}.

\subsection{Sample preparation}

In the experiment we use milli-Q demineralised water in a $25$~mL beaker. This beaker is placed inside a container in which the microscope objective is introduced through a closely fitting hole in the lid that is sealed by using rubber O-rings (see figure 1 (b)). The isolated air inside is then allowed to saturate with water vapor for a few minutes, to limit the speed with which the water level lowers due to evaporation to less than $1~\mu$m/hour, reducing the need to periodically refocus the surface of the target. The gaseous atmosphere of the cuvette can be controlled using a circulating system with a supply and an exhaust. For the experiments we use three different gasses, namely Helium, air and 1,1,1,2 tetrafluoroethane. Table I shows relevant physical properties of the gases.

\begin{center}
$\bf{Table I}$: Relevant gas properties at standard temperature and pressure. 
	\label{Table I}
	\begin{tabular}{ | l | l| l | p{2.9cm} |}
    \hline & Molar Mass (g/mol) & Density (mg/cm$^ {3}$) & Speed of sound (m/s) \\
    \hline 1,1,1,2 Tetrafluoroethane & 102.03 & 4.25 & 168.41 \\ \hline
    Air & 18.00 & 1.20 & 343.21 \\ \hline
    Helium & 4.00 & 0.18 & 972.00 \\ \hline
    \end{tabular}
\end{center}

\begin{figure}[ht]
\includegraphics[width=13cm]{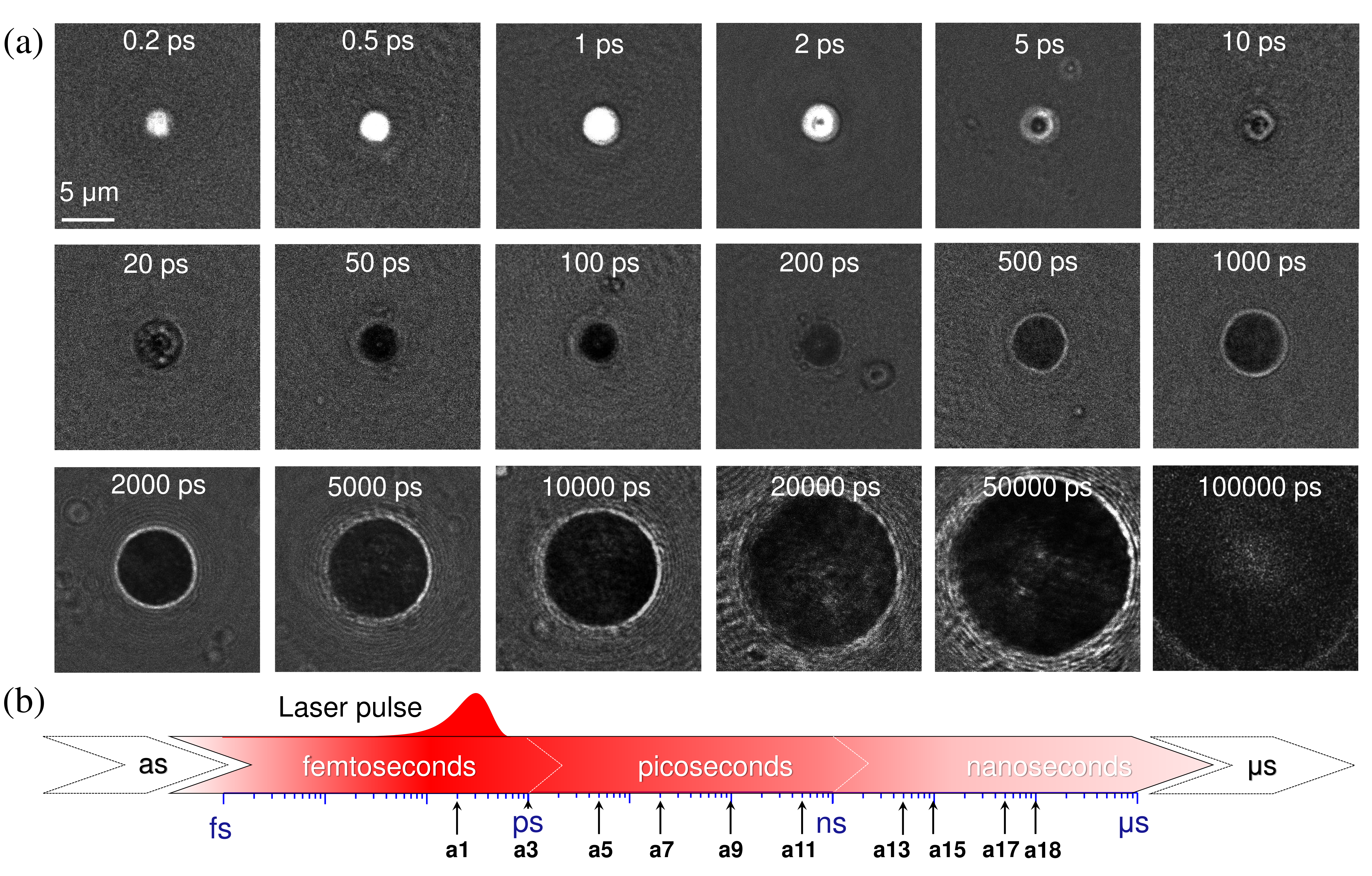}
\caption{(a) Typical transient differential reflectivity maps of the water/air interface obtained for a pump pulse fluence of 25~J/cm$^{2}$. The images share the same lateral scale (20x20~$\mu$m$^2$) and are represented using the same color scale (which saturates the first four images). In the first few picoseconds, we see a strong increase of the reflectivity. After 2 picoseconds, we see the increase in reflectivity turn into a ring, which fades for longer pump-probe delays. Starting at approximately 10 picoseconds, we see a reduced differential reflectivity starting in the center. On the timescale of 100 ps, the dark region radially expands and a bright rim develops around it. (b) Time-line during the ablation of water. The time-delays, at which the images in (a) are taken, are indicated by the arrows labelled a1 to a18.}
\label{fig2}
\end{figure}

\section{Results}

\subsection{Ultrafast ablation of water through the time-scales}
Using the experimental setup, we investigate the transient reflectivity of a water/gas interface during ultrafast laser ablation of water upon tight focusing conditions, from 10 fs to 100 ns. Typical examples of such images, measured in an air atmosphere, are shown in Fig.~\ref{fig2} (a) whose time delays are chosen to be equally spaced in a logarithmic scale, as presented in Fig.~\ref{fig2} (b). Here, three main regimes are observed: excitation and relaxation of a dense electron plasma (first row), evaporation onset (second row) and water vapor expansion (third row). We observe a delay of few tens of picoseconds before the laser-induced water vapor noticeably starts to expand laterally, which agrees with the expansion dynamics observed inside bulk water \cite{Schaffer2002} and at the surface of a water jet \cite{Sarpe-Tudoran2006} under similar focusing conditions.

\begin{figure}[ht]
\includegraphics[width=13 cm]{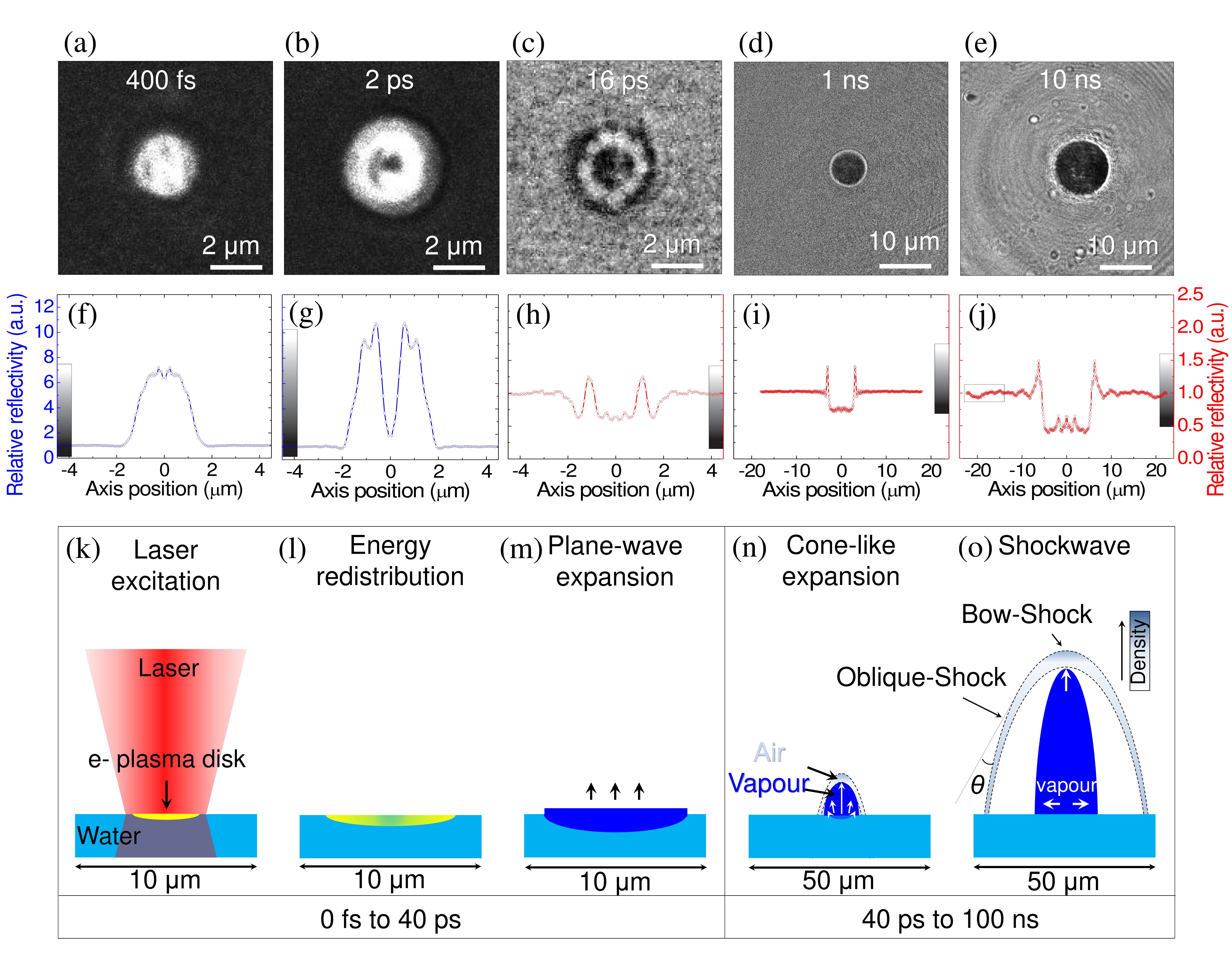}
\caption{(a)-(e) Transient relative reflectivity maps representative of the different time regimes. (f)-(j) Radial average of the above images. The blue curves share the vertical scale of (f), the red curves share the vertical scale of (j). The color-bar indicating the grey-scale of the images (a)-(e) is attached to the vertical axes of the graphs (f)-(j). (k)-(o) Schematics of the physical processes, where (f)-(h) extends from 0 fs to 40 ps, (i)-(j) spreads from 40 ps to 100 ns. Note that for a given column the images, profiles and schematics have the same lateral scale.}
\label{fig3}
\end{figure} 

In Figs.~\ref{fig3} (a)-(e), we show images representative of the different time-separated regimes, combined with their radial averages (f)-(j) and schematics to discuss the corresponding physical processes (k)-(o). Initially the energy of the laser is coupled into the irradiated water via strong field ionization \cite{keldysh1965ionization}, $\it{i.e.}$ multiphoton and tunneling ionization. Afterwards, seed electrons that are excited via non-linear ionization processes gain kinetic energy by means of inverse bremsstrahlung leading to impact ionization\cite{Sarpe-Tudoran2006, Linz2015, Rethfeld2006}. Consequently, during the first picosecond, we observe a strong increase of the surface reflectivity that we atribute to the formation of a well-localized dense electron plasma, as illustrated in Fig.~\ref{fig3} (k). The maximum reflectivity increase that we observe experimentally is $\Delta$R = 0.16, whereas the reflectivity of unexcited water is R$_{o}$ = 0.02. According to the Drude model for a free electron gas this corresponds to an electron density on the order of 10$^{22}$ $cm^{-3}$. These findings are consistent with the transient reflectivity contrast reported for solid crystalline and glassy dielectric targets \cite{Hernandez-Rueda2012, Hernandez-Rueda2015}, whose electonic and linear optical properties are comparable to those of water. After 2 picoseconds, the reflectivity in the center starts to drop. We attribute this reduction in reflectivity to the fact that the water in the center starts to evaporate, locally reducing the density of both the water and the electron plasma, as illustrated in Fig.~\ref{fig3} (l). We therefore use the appearance of this reduction to estimate the ablation threshold to be 8.1~J/cm$^{2}$, considering the calibrated 1/e$^2$ beam waist (see previous section and \cite{liu1982simple}) and the Gaussian distribution of the laser. Curiously, we reproducibly observe a periodic azimuthal structure in the ring of increased reflectivity for delays that range from 8 ps to 40 ps, as shown in Fig.\ref{fig3} (c). The orientation of this periodic structure remains unaltered for different irradiation experiments at a given time-delay, indicating a deterministic behaviour. As of yet, we have no explanation of the cause of this structure. As time goes on, water further away from the center also evaporates, expanding upwards, as illustrated in Fig.\ref{fig3} (m), causing the corrugated ring of increased reflectivity to fade away for longer pump-probe delays. Starting at approximately 10 picoseconds, we see an overall reduced differential reflectivity in the affected area. We attribute this to the fact that the hot water vapor above the water/air interface can act as a local anti-reflection coating that absorbs the light of the probe beam \cite{Schaffer2002, Sarpe-Tudoran2006}. From 100 ps onwards, we observe that the area of reduced reflectivity radially expands and develops a sharp bright rim. We can understand that this happens when the expanding vapor cloud becomes larger in radius than the initially laser-excited area, as illustrated in Fig.\ref{fig3} (n). The bright edge in Figs.\ref{fig3} (d),(e),(i),(j) can be interpreted as the projection of the contact discontinuity between the supersonically expanding vapor and the surrounding air. The fact that the contact discontinuity is sharply visible means it must be widest near the focal plane, $\it{i.e.}$ the water/gas interface. This strongly suggests that the expansion is cone-like, as illustrated in Fig.\ref{fig3} (n),(o), which in turn indicates a highly supersonic expansion. This is supported by the fact that we observe a shockwave in the surrounding air, as can be seen in Figs.~\ref{fig3} (e),(j).

\subsection{Atmosphere influence during ultrafast ablation of water}

\begin{figure}[ht]

\includegraphics[width=14cm]{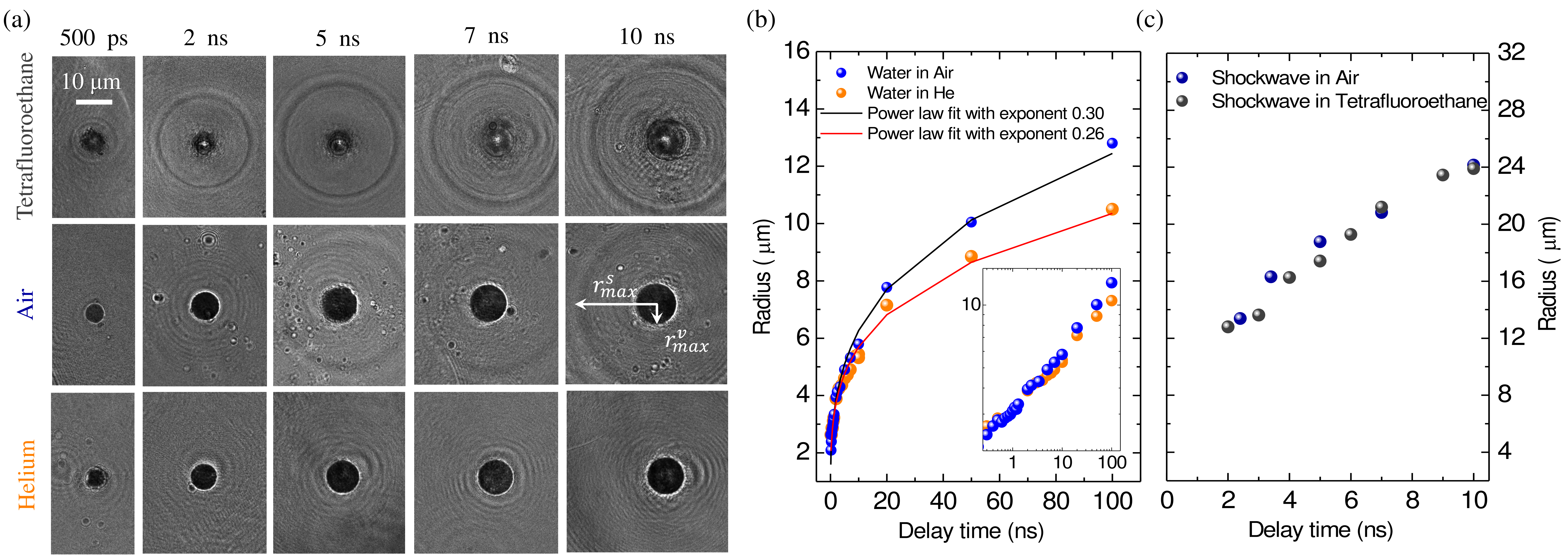}
\caption{(a) Representative snapshots of the ablation process during vapor expansion in the presence of tetrafluoroethane, air and Helium. We use the same laser fluence employed in section 3.1. (b) Radius of the front bright rim as a function of pump-probe delay for experiments carried out in a water/air (dark blue) and a water/He (orange) interfaces. Each experiment averages the radius of 20 images. The inset shows the data in a log-log scale. (c) Radius of the shockwave in air (dark blue) and tetrafluoroethane (black) as function of time.}
\label{fig4}
\end{figure}

The expansion of the water vapor also depends on the atmosphere that surrounds the irradiated area. Fig.~\ref{fig4} (a) presents snapshots of the transient reflectivity during the ablation of water in the presence of different gases, from 500 ps to 10 ns. As explained in Fig.~\ref{fig3} the dark disk in the center ($r_{\rm{max}}^{\rm{v}} \approx 5.5~\mu$m) is attributed to the rapidly expanding water vapor, whereas the outermost dark and bright ring ($r_{\rm{max}}^{\rm{s}} \approx 24~\mu$m) corresponds to the shockwave propelled in the surrounding gas (see white arrows in Fig.~\ref{fig4} (a) ''air'' at 10 ns). The increment in the lateral size as well as the sharpness of the vapor front are clearly influenced by the surrounding atmosphere. In the presence of Helium and Air, the vapor resembles a homogeneous dark circle with a distinct edge, while in a tetrafluoroethane atmosphere, the dark area is heterogeneous and lacks a sharp edge. Although the images are similar in appearance, Fig.~\ref{fig4} (b) shows that the radial expansion in air achieves a slightly larger radius than in He for long delays up to 100 ns. For the first two nanoseconds, the same expansion velocity is observed in the presence of both gases ($\approx$ 810 m/s), which corresponds to a supersonic expansion for air ($v_{\rm{s}}$ = 343 m/s, $v_{\rm{s}}$ is the speed of sound) and a subsonic expansion for Helium ($v_{\rm{s}}$ = 972 m/s). This initial expansion speed explains why a shockwave is experimentally observed in an air atmosphere but not in a Helium environment. After 2 ns, the radial increase slows down more rapidly in a He environment than in the air environment. We use a Sedov-Taylor formula to estimate the energy released during the expansion of the water vapor in the presence of air and He (see Fig.~\ref{fig4} (b))

\begin{equation}
r(t,E)=\bigg(\frac{3E}{\pi\rho}\bigg)^{1/5}t^{2/5} \\
\end{equation}

where $\rho$ is the density of water and E stands for the energy. Considering a hemispherical expansion, we estimate the energies released in an air and He atmospheres to be 618 pJ and 360 pJ, respectively. To test the validity of using the Sedov-Taylor model, we additionaly fit a power-law formula $r~\propto~t^{\alpha}$, retrieving an exponent of 0.30 and 0.26 for air and Helium, respectively, as shown in Fig.~\ref{fig4} (b). This shows that the expansion is not fully self-similar. This lack of self-similarity can be qualitatively understood, as we expect that in both atmospheres, the superheated water initially undergoes a vertical supersonic expansion \cite{Arnold1999, vogel2005mechanisms}. During the expansion, the expanding vapor gathers mass by sweeping up the gas from the atmosphere, eventually leading to a transition from a vertical expansion to a more isotropic expansion. In an atmosphere with a low mass density ($\rho _{\rm{air}}$~$\approx~7\rho _{\rm{He}}$), this transition will occur later. Thus the system will favour vertical expansion over lateral expansion as compared to a system with an atmosphere with a high mass density, in correspondence with the results shown in Fig.~\ref{fig4} (b). 

Fig.~\ref{fig4} (a) also shows that a gaseous atmosphere with higher density ($\rho_{\rm{TFE}} \approx 4\cdot\rho_{\rm{air}}~\approx 24\cdot\rho_{\rm{He}}$, see table I in the experimental section) presents a shock with higher optical visibility, $\it{i.e.}$ tetrafluoroethane. The graph in Fig.~\ref{fig4} (c) shows that the shock radius as a function of time has a similar behaviour in both atmospheres.

\section{Conclusion}

In conclusion, we experimentally study the ultrafast laser ablation at a water/gas interface by using time-resolved microscopy. We present the changes in the transient reflectivity for several time-scales, from 10 fs to 100 ns. Overall, our work explores the initial laser-water interaction (femtoseconds), the intermediate extreme thermodynamic state of the system (picoseconds) and the subsequent compressible fluid dynamics (nanoseconds). As an outlook we propose to further link and explore the long-run mechanical behaviour, $\it{i.e.}$ surface waves, which will require delays far into the microsecond regime. We additionally test the influence of the atmosphere during the ablation dynamics, finding that the propulsion of a shock in the surrounding gas depends on its speed of sound and molecular weight. Moreover, we find that the atmosphere influences the way the laser induced vapor expands during the first 100 ns, following a Sedov-Taylor power law expansion with different exponents, which indicates the process is not self similar. The understanding of this concatenation of physical processes has high impact and tremendous potential in the field of laser nano-surgery ($\it{i.e.}$ cell, ocular and neuro surgery) \cite{ vogel2005mechanisms, Tirlapur2002}, since they are behind the behaviour of photomechanical damage. As the ablation of water involves less phase transitions than the ablation of solid samples, our study can contribute further research on the ultrafast laser ablation of more complex systems, such as nanoparticle synthesis via ablation of inmersed targets \cite{mafune2000formation} or EUV-light generation for nanolithograpy via tin droplet ablation \cite{wagner2010euv}.

\section*{Funding}

European Commission, Horizon 2020, Marie Skłodowska-Curie Action Individual Fellowship (703696 ADMEP).

\section*{Acknowledgments}

The authors thank Denise Krol, Hanneke Gelderblom, Allard Mosk and Ingmar Swart for fruitful discussions. The authors also thank Paul Jurrius, Cees de Kok and Dante Killian for discussions and technical assistance.

\end{document}